\documentclass[fleqn,usenatbib]{mnras}

\usepackage{newtxtext,newtxmath}
\usepackage{todonotes}

\usepackage[T1]{fontenc}

\DeclareRobustCommand{\VAN}[3]{#2}
\let\VANthebibliography\thebibliography
\def\thebibliography{\DeclareRobustCommand{\VAN}[3]{##3}\VANthebibliography}

\usepackage{graphicx}	
\usepackage{amsmath}	
\usepackage{xcolor}

\newcommand{\lambdaint}{\lambda_{\textrm{int}}}
\newcommand{\kappaext}{\kappa_{\textrm{ext}}}
\newcommand{\DsDds}{D_{\textrm{s}}/D_{\textrm{ds}}}
\newcommand{\Dd}{D_{\textrm{d}}}
\newcommand{\Ds}{D_{\textrm{s}}}
\newcommand{\Dds}{D_{\textrm{ds}}}
\newcommand{\Ddt}{D_{\Delta\textrm{t}}}

\newcommand{\sref}[1]{Section~\ref{#1}}
\newcommand{\fref}[1]{Figure~\ref{#1}}
\newcommand{\tref}[1]{Table~\ref{#1}}
\newcommand{\eref}[1]{Equation~(\ref{#1})}
\newcommand{\aref}[1]{Appendix~(\ref{#1})}

\title[AO PSF-R: Meeting the astrometric requirements for TDC]{Point-spread function reconstruction of adaptive-optics imaging: Meeting the astrometric requirements for time-delay cosmography}

\author[G.~C.-F.~Chen et al.]{
Geoff~C.-F.~Chen,$^{1}$\thanks{E-mail: gcfchen@astro.ucla.edu}
Tommaso Treu,$^{1}$
Christopher~D.~Fassnacht,$^{2}$
Sam Ragland,$^{3}$
Thomas Schmidt$^{1}$
\newauthor{
and Sherry H. Suyu$^{4,5,6}$}
\\
$^{1}$Physics and Astronomy Department, University of California, Los Angeles, CA 90095, USA\\
$^{2}$Department of Physics and Astronomy, University of California, Davis, CA 95616, USA\\
$^{3}$W. M. Keck Observatory\\
$^{4}$Max Planck Institute for Astrophysics, Karl-Schwarzschild-Str. 1, 85748 Garching, Germany\\
$^{5}$Physik-Department, Technische Universit\"at M\"unchen, James-Franck-Str. 1, 85748 Garching, Germany\\
$^{6}$Institute of Astronomy and Astrophysics, Academia Sinica, 11F of ASMAB, No.1, Section 4, Roosevelt Road, Taipei 10617, Taiwan
}

\date{Accepted XXX. Received YYY; in original form ZZZ}

\pubyear{2015}

\begin{document}
\label{firstpage}
\pagerange{\pageref{firstpage}--\pageref{lastpage}}
\maketitle

\begin{abstract}
Astrometric precision and knowledge of the point spread function are key ingredients for a wide range of astrophysical studies including time-delay cosmography in which strongly lensed quasar systems are used to determine the Hubble constant and other cosmological parameters. Astrometric uncertainty on the positions of the multiply-imaged point sources contributes to the overall uncertainty in inferred distances and therefore the Hubble constant. Similarly, knowledge of the wings of the points spread function (PSF) is necessary to disentangle light from the background sources and the foreground deflector. We analyze adaptive optics (AO) images of the strong lens system J\,0659+1629 obtained with the W. M. Keck Observatory using the laser guide star AO system. We show that by using a reconstructed point spread function we can i) obtain astrometric precision of $< 1$  mas, which is more than sufficient for time-delay cosmography; and ii) subtract all point-like images resulting in residuals consistent with the noise level. The method we have developed is not limited to strong lensing, and is generally applicable to a wide range of scientific cases that have multiple point sources nearby.
\end{abstract}




\section{Introduction}
Strong gravitational lensing time delays provide a one-step measurement of cosmological distances in the
Universe \citep{Refsdal64}. Hence they can be used to determine the Hubble constant independent of the traditional distance ladder method \citep{RiessEtal19}.
In a time-delay lens, the lensed background is composed of
a time variable point-like source, usually an active galactic nucleus (AGN) or a supernova, and its host galaxy. The time delays between the images of the lensed source, induced by the
foreground lens, are given by 
$\Delta t$ $=\frac{1}{c}D_{\Delta t}\Delta\tau$.
Here, the time-delay distance, $D_{\Delta t}$, depends on
cosmological parameters, in particular the Hubble constant, $H_0$
\citep[e.g.,][]{SuyuEtal10}, whereas $\Delta\tau$ represents the gravitational potential difference between image positions, which depends on the geometry of the lens system.
The gravitational potential of the foreground lens galaxy, $\tau$, can be constrained by the spatial extent of the lensed 
background galaxy (usually known as ``arcs'') \citep[e.g.,][]{KochanekEtal01, SuyuEtal09}, combined with 
stellar kinematics of the lens
\citep[e.g.,][]{TreuKoopmans02, KoopmansEtal03b, SuyuEtal10, SuyuEtal14, YidrimEtal20}
and studies of the lens environment that are performed through numerical ray-tracing simulations \citep[e.g.,][]{HilbertEtal07, HilbertEtal09, SuyuEtal10, FassnachtEtal11, GreeneEtal13, CollettEtal13, RusuEtal17,RusuEtal20_H0LiCOW} or weak lensing \citep{TihhonovaEtal18,TihhonovaEtal20}.

Therefore, by measuring the time delays
between the multiple images and modeling the lens system as well as the relevant line-of-sight mass 
distribution, we can infer $D_{\Delta t}$.
Furthermore, the time delays -- in combination with
stellar velocity dispersion measurements of the lens galaxy -- allow us to infer the angular diameter distance ($\Dd$) to the lens galaxy
\citep{ParaficzHjorth09, JeeEtal15}, thereby providing additional cosmological information.

From a technical point of view, a key ingredient to successful time-delay cosmography is knowledge of the point spread function (PSF) in the high-resolution imaging that is used to constrain the mass model. The PSF is needed to derive precise astrometry of the multiply-imaged variable point source \citep{BirrerTreu19} and to disentangle in the image the light from the quasar's host galaxy light and the contribution from the foreground deflector.

Most of the work on time delay cosmography in the past two decades has been done using Hubble Space Telescope data, exploiting its sharp and stable point spread function. However, adaptive optics (AO) technology, a technique to improve the performance of optical/near infrared systems by reducing the effect of incoming wavefront distortions \citep[e.g.,][]{RoussetEtal90,Beckers93,Watson97,wizinowich06}, has improved substantially over recent years, making it possible to obtain high-resolution images from ground based telescopes that can be used for time-delay cosmography \citep{GChenEtal16,GChenEtal20}.

The key problem is that the AO PSF varies temporally and spatially, therefore the PSF needs to be reconstructed, either from the data itself or from telemetry data acquired during the observations. \citet{GChenEtal16} solved this problem by reconstructing the PSF from using only the data. They exploited the information provided by the multiple lensed quasar images, showing that one can reach mass-model precision comparable, or superior, to the HST PSF especially for intrinsically red sources where the AO system performs best. \citep{GChenEtal19}.  

In this paper, we continue our investigation of the precision and accuracy of time delay cosmography with adaptive optics, by examining the astrometric error budget of multiply-imaged quasars. In addition to the PSF reconstructed from the data, as proposed by \citet{GChenEtal16}, hereafter PSF-CS, we also consider a PSF reconstructed from telemetry data, PSF-R. We show that the two methods are highly complementary, with PSF-CS providing the best performance near the core of the PSF, while PSF-R provides the most information in the wings. By combining these two approaches we show that sub-mas astrometric precision can be achieved and that PSF-CS+R can be subtracted from the data, leaving residuals consistent with the level of noise.

The paper is organized as follows: In~\sref{sec:tdcf} we briefly review the time delay formalism to set the notation. In~\sref{sec:astrometry} we extend the formalism introduced by \citet{BirrerTreu19} to compute the contribution of the astrometric error budget to the main deflector's distance, $\Dd$. In~\sref{sec:AOreduction} we present the AO data, while in~\sref{sec:PSF} we compare the performance of the different PSF reconstruction methods. In~\sref{sec:H0_AOHST} we compare the performance of AO and HST astrometry with the the requirements to derive $H_0$, followed by a brief summary that concludes the paper in~\sref{sec:summary}.

\section{Time-delay cosmography formalism}
\label{sec:tdcf}
When a light ray passes near a massive object, its trajectory is deflected by the gravitational potential of the so-called deflector, resulting in a time delay compared to the travel time absent the deflector. The excess time delay is given by
\begin{equation}
\label{eq:theory6}
t(\theta, \beta)=(1-\kappaext)\lambdaint\frac{D_{\Delta t}}{c}\left[\frac{1}{2}(\theta-\beta)^{2}-\psi(\theta)\right],
\end{equation}
where $\theta$, $\beta$ are the image position and the source position, respectively, while $\psi(\theta)$ represents the gravitational potential of the lens at point $\theta$. The two parameters, $\kappaext$ and $\lambdaint$, are related to the mass-sheet transformation (MST) \citep{FalcoEtal85,GorensteinEtal88,FassnachtEtal02,SuyuEtal13,GreeneEtal13,CollettEtal13,Kochanek20,Kochanek21,BirrerEtal20,GChenEtal20}. Specifically, $\kappaext$ represents the external MST, which is associated with mass along the line of sight and $\lambdaint$ represents the internal MST, associated with transformation of the deflector's mass profile \citep{GChenEtal20}. The angular term in brackets in \eref{eq:theory6} is called the Fermat potential, $\phi(\theta,\beta)$.  

The relative time delay measured between image A and image B can be expressed as
\begin{equation}
    \Delta t_{\rm AB} =\frac{\Ddt}{c}\Delta\phi_{\rm AB}.
\end{equation}
The time-delay distance is defined as
\begin{equation}
\label{eq:theory7}
\Ddt\equiv(1+\textit{z}_{\rm d})\frac{D_{\rm d}D_{\rm s}}{{D_{\rm ds}}}\propto H_{0}^{-1},
\end{equation}
$D_{\rm d}$,
$D_{\rm s}$ and $D_{\rm ds}$ are the angular diameter distances to
the lens, to the source, and between the lens and the source,
respectively, and $z_{\rm{d}}$ represents the main deflector's redshift.
Following Fermat's principle, the gradient of the excess time delay given by 
equation (\ref{eq:theory6}) vanishes at the position of the lensed images, which yields the so-called lens equation
\begin{equation}
\beta = \theta -
\nabla\psi(\theta),
\end{equation}
that governs the deflection of light rays in the thin lens approximation.

Under the MST, the dependence of projected stellar velocity dispersion on the mass model, $\sigma_v^{\textrm{p}}$, can be written as  
\citep{GChenEtal20}
\begin{equation}
\label{eq:vd_nolambda}
    (\sigma_v^{\textrm{p}})^{2}=(1-\kappaext)\lambdaint\left(\frac{\Ds}{\Dds}\right) c^{2}J(\eta_{\textrm{lens}},\eta_{\textrm{light}},\beta_{\textrm{ani}}),
\end{equation}
where $J$ contains the angular-dependent information in the lens modeling and the stellar orbital anisotropy distribution, $\beta_{\textrm{ani}}$ \citep[see details in][]{JeeEtal15}.
We can replace the MST-related terms ($\lambdaint$ and $\kappaext$) with \eref{eq:vd_nolambda} and the predicted time delays will directly relate to the velocity dispersion via
\begin{equation}
\label{eq:Dd}
    \Delta t_{\rm AB}=(1+z_{\rm d})\frac{\Dd }{c}\frac{\Delta\phi_{\rm AB}(\theta, \beta)}{ J(\eta_{\textrm{lens}},\eta_{\textrm{light}},\beta_{\textrm{ani}})}\frac{(\sigma_v^{\rm{p}})^2}{c^{2}}.
\end{equation} 
%
Once the time delay and velocity dispersion are measured, the value of $\Dd$ can be determined. 
When further including information about the environment (which provides an estimate for $\kappaext$) and $\DsDds$ information which comes from additional data such as spatially resolved kinematics, external datasets, or the assumption of a cosmological model, one can determine $\lambdaint$ \citep[][Yildirim et al. in prep.]{GChenEtal20,BirrerEtal20}.

\section{Astrometric error propagation of $H_{0}$}
\label{sec:astrometry}

The astrometric uncertainty on the lensed quasar positions can affect the estimation of the relative Fermat potential, $\Delta\phi_{\rm AB}$, no matter how precisely the lensing potential is determined from the imaging \citep{BirrerTreu19}. Therefore, the astrometric precision can affect the determination of $\Dd$ and $\Ddt$ and hence $H_{0}$. 
\citet{BirrerTreu19} have described the error propagation to $H_{0}$ given a $\Ddt$ measurement.
They show that, if the $H_{0}$ information comes from $\Ddt$, the Hubble constant scales as 
\begin{equation}
\label{eq:H0_error_Ddt}
    \delta H_{0}\sim\frac{\delta\beta}{\theta_{\rm AB}},
\end{equation} 
where $\theta_{\rm AB}$ is the image separations between imaging A and B.
We extend here the formalism introduced by \citet{BirrerTreu19} to compute the contribution of the astrometric error budget to the $\Dd$ distance.
Given \eref{eq:Dd}, we can express the error propagation of velocity dispersion, $\sigma_{v}$, relative Fermat potential, $\Delta\phi_{\rm AB}$, and time delays, $\Delta t_{\rm AB}$, to $H_{0}$ as
\begin{equation}
    \frac{\delta H_{0}}{H_{0}}\sim - \frac{\delta \Dd}{\Dd}\sim \sqrt{4\frac{\delta \sigma_v^2}{\sigma_v^2}+\frac{\delta \Delta t^2_{\rm AB}}{\Delta t^2_{\rm AB}}+\frac{\delta \Delta \phi^2_{\rm AB}}{\Delta \phi^2_{\rm AB}}},
\end{equation}
where
\begin{equation}
    \frac{\delta \Delta \phi_{\rm AB}}{\Delta \phi_{\rm AB}}=(1+z_{\rm d})\frac{\Dd }{c}\frac{\sigma_v^2}{c^{2}}\frac{\theta_{\rm AB}}{ J\Delta t_{\rm AB}}\delta \beta,
\end{equation}
%

In order for the astrometric uncertainty to be subdominant with respect to the uncertainty in the time delay measurement, $\Delta t_{\rm AB}$, the following requirement applies
\begin{equation}
    (1+z_{\rm d})\frac{\Dd }{c}\frac{\sigma_v^2}{c^{2}}\frac{\theta_{\rm AB}}{ J\Delta t_{\rm AB}}\delta \beta<\frac{\delta \Delta t_{\rm AB}}{\Delta t_{\rm AB}}.
\end{equation}
Since $\sigma_{v}^{2}\propto\theta_{\rm AB}$, $\Delta t_{\rm AB}\propto\theta_{\rm AB}^{2}$, and $J\propto\theta_{\rm AB}$, the Hubble constant uncertainty due to astrometric error scales as 
\begin{equation}
    \delta H_{0}\sim\frac{\delta\beta}{\theta_{\rm AB}},
\end{equation}
which is the same as \eref{eq:H0_error_Ddt}.

In conclusion, the requirements for $\Dd$ are the same as for $\Ddt$. The most important effect is that the required astrometric precision, at fixed $H_0$ precision, scales inversely with the image separation.


\section{Keck Adaptive Optics Imaging}
\label{sec:AOreduction}
The AO imaging of J\,0659+1629 was obtained at K$^\prime$-band with the Near-infrared Camera 2 (NIRC2) during an engineering night on December 02, 2020. The target was observed with the narrow camera setup, which provides a roughly 10$\times$10\arcsec\ field of view and a pixel scale of 9.942 milliarcsec (mas). 
The total exposure time was 1440 seconds.
We follow our previous work \citep{GChenEtal16,GChenEtal19} and use the SHARP python-based pipeline, which performs a flat-field correction, sky subtraction, correction of the optical distortion in the images, and a coaddition of the exposures. 
For the distortion correction step, the images are resampled to produce final pixel scales of 10~mas pix$^{-1}$ for the narrow camera. The narrow camera pixels well sample the AO PSF, which has typical FWHM values of 60--90~mas.  
To improve the modeling efficiency for the narrow camera data, we perform a 2$\times$2 binning of the images produced by the pipeline to obtain images that have a 20~mas pix$^{-1}$ scale. We note that at this scale the PSF is adequately sampled, with 3-4 pixels per FWHM.

\begin{figure*}
    \centering
    \includegraphics[width=0.85\linewidth]{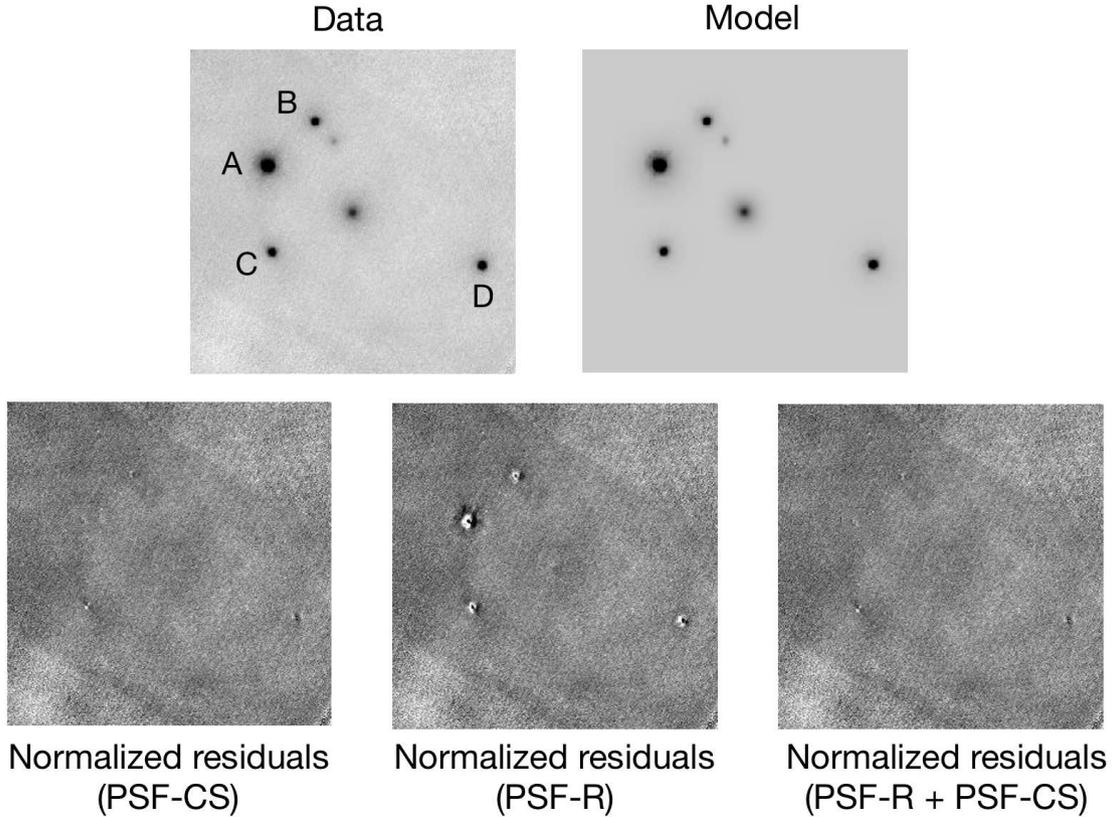}
    \caption{Imaging data, model, and normalized residuals given PSF-CS, PSF-R, and PSF-R+PSF-CS. The model image shown above is created by PSF-CS. 
    For all three systems, once the PSF-CS is applied, we can model the lensed quasar down to the noise level. PSF-CS utilizes multiple concentric gaussians as the initial PSF model and performs an iterative corrections on the PSF model given the residuals \citep[see more details in][]{GChenEtal16}.}
    \label{fig:J0659_residuals}
\end{figure*}



\section{Comparison of the reconstructed PSF from different methods}
\label{sec:PSF}
Accurate knowledge of the PSF structure is the key ingredient for time-delay cosmography. 
We use the AO imaging of J\,0659+1629 to examine the performance of the state-of-the-art PSF reconstructed methods. 
In \sref{subsec:residuals}, we show the residuals given different PSF models. In \sref{subsec:structures}, we compare the reconstructed PSF structures including the core and wing. In \sref{subsec:astrometric}
we compare the lensed quasar positions inferred from different PSFs, and compare the astrometric precision between AO and HST.
\subsection{Residuals}
\label{subsec:residuals}
Modeling the AO imaging of lensed quasar system down to the noise level requires accurate description of the AO PSF structures which can vary significantly given different observational conditions. 
To solve this problem, one can reconstruct the PSF either from the data themselves by exploiting the information provided by the multiple lensed quasar images \citep[hereafter PSF-CS;][]{GChenEtal16} or from telemetry data acquired during the observations 
(hereafter PSF-R; Ragland 2018, Ragland et al. 2018)
\citep[hereafter PSF-R;][]{Ragland18,RaglandEtal18}. 
The errors from the PSF-R are listed in the \aref{appexdix1}.

In \fref{fig:J0659_residuals}, we show the imaging data, the model (lensing galaxy light and lensed quasars), and the normalized residuals given PSF-R, PSF-CS and PSF-R+PSF-CS, where PSF-R+PSR-CS means that we use PSF-R as an initial PSF and then perform PSF corrections on it. PSF-CS use concentric multiple gaussians as the intial PSF and perform the PSF corrections.
While there are some residuals in the center of the lensed quasars images using PSF-R,
we can build accurate PSF structures and model the data down to the noise level, except a very small region right in the center of the PSF core, once we apply the PSF-CS method by exploiting the fact that the four lensed quasar images share the same PSF structure.

\begin{figure*}
    \centering
    \includegraphics[width=0.7\linewidth]{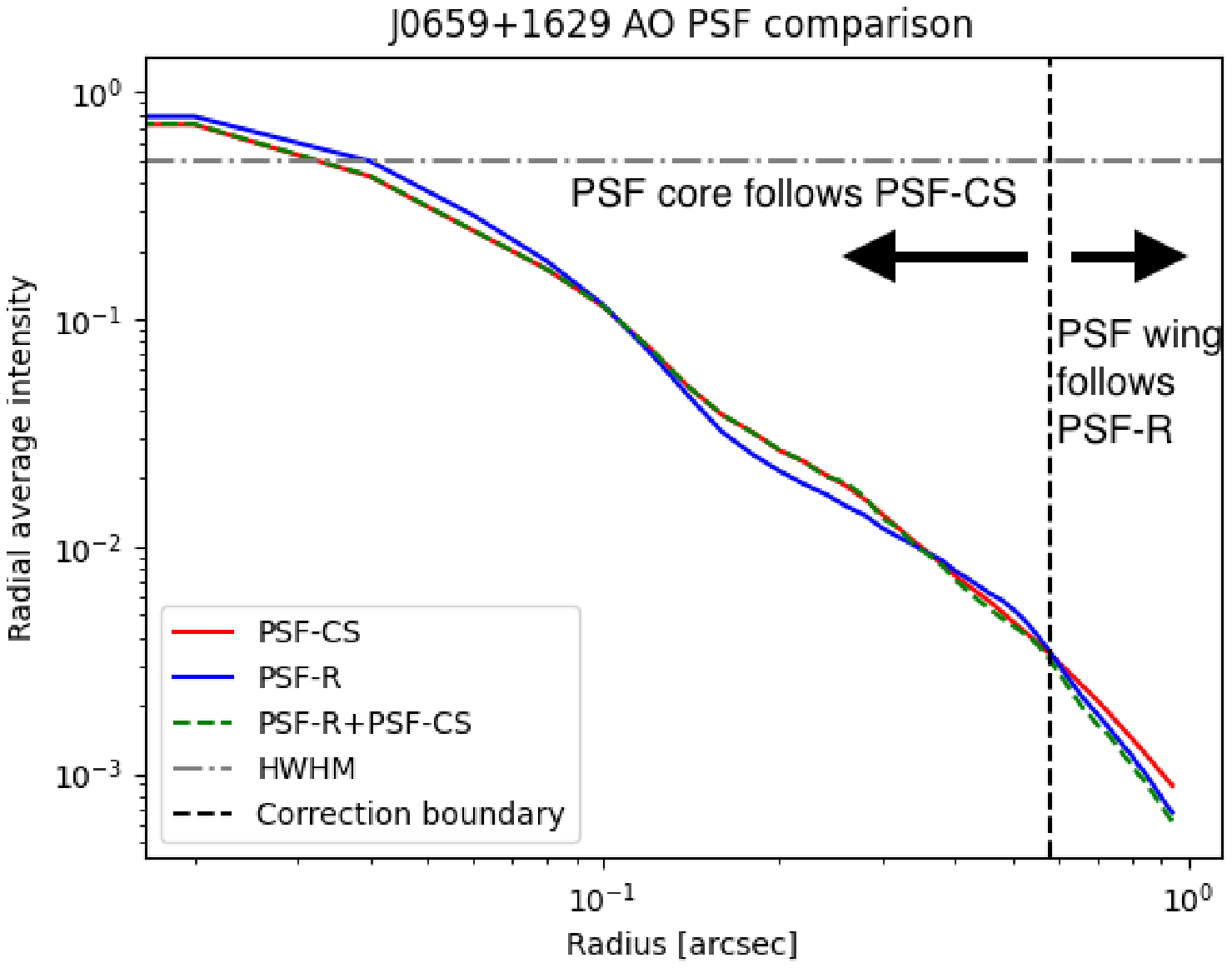}
    \caption{Comparison of the azimuthally-average radial profile of the reconstructed PSFs from PSF-CS \citep{GChenEtal16}, PSF-R, and PSF-R + PSF-CS. All three PSF models are normalised by the central brightest pixel.
    One can can see that inside the correction boundary, PSF-R+PSF-CS follows PSF-CS, while outside the correction boundary PSF-R+PSF-CS follows PSF-R. In other words, PSF-CS can provide an accurate description of the true PSF model in the center of the lensed quasar, while PSF-R can provide the information relevant to the wings \citep{GChenEtal16}.}
    \label{fig:J0659_structure}
\end{figure*}

\subsection{PSF wing structures: cores and wings}
\label{subsec:structures}

A typical AO PSF consists of a roughly diffraction-limited core and extended wing structure. Although the PSF-CS approach allows one to model the AO imaging down to the noise level, \citet{GChenEtal19} found that there is a degeneracy between the AO PSF wing and the lens light if the lensing galaxy is very extended. This can potentially bias the inference of the baryonic matter distribution if one uses the lens light as the tracer for the baryonic matter. It is thus important to characterize the PSF wing from external information.

In \fref{fig:J0659_structure}, we show the comparison of the azimuthally average intensity of the reconstructed PSFs from PSF-CS, PSF-R, and PSF-R +PSF-CS.
We can see that once the correction is applied on PSF-R, the green dashed line (PSF-R +PSF-CS) agrees with the red line (PSF-CS). The correction makes the residuals of PSF-R + PSF-CS in \fref{fig:J0659_residuals} disappear. This indicates that the core structures can be well determined. However, although the red line (PSF-CS) and green line (PSF-R +PSF-CS) have different PSF wings, the residuals are indistinguishable. Therefore, the flexibility of the PSF wings could potentially introduce surface brightness degeneracy with the lens light.


With the wing information from PSF-R, we show that outside the correction boundary, PSF-R+PSF-CS follows PSF-R and hence PSF-R can be used to break the degeneracy.

\subsection{Astrometric precision}
\label{subsec:astrometric}
\begin{figure*}
    \centering
    \includegraphics[width=0.9\linewidth]{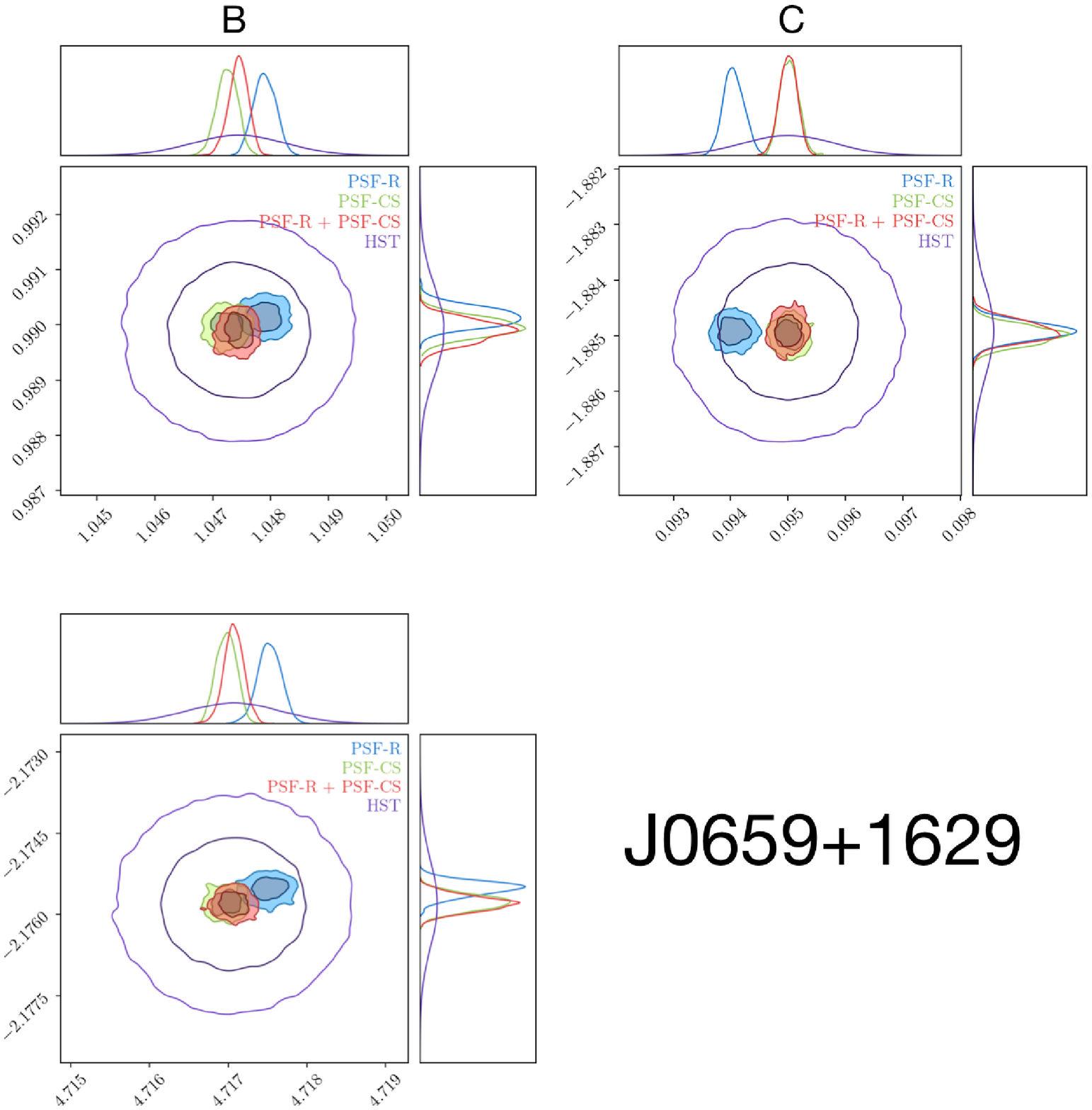}
    \caption{Comparison of the lensed quasar positions inferred by using PSF-R, PSF-CS, and PSF-R+PSF-CS, and of the precision obtained with AO vs. HST. The relative positions (in arcsecond) of lensed images B, C, and D are relative to the lensed image A. The residuals given PSF-R shown in \fref{fig:J0659_residuals} can affect the determination of the lensed quasar in 1 to 2 mas level (blue v.s. green and red), but after applying the correction to the PSF-R, the lensed quasar positions agree with the results from PSF-CS. The results indicate that the position of the lensed quasars can be robustly determined well below mili-arcsecond level.}
    \label{fig:J0659_precision}
\end{figure*}

Astrometric uncertainty can contribute to total error budget of the distance measurements and hence $H_{0}$. We compare the lensed quasar positions inferred by using PSF-R, PSF-CS, and PSF-R+PSF-CS. We also compare the precision obtained with AO and HST. We show the relative positions of lensed quasar images B, C, an D with respect to the lensed quasar image A in \fref{fig:J0659_precision}. One can see that the residuals given PSF-R shown in \fref{fig:J0659_residuals} can affect the determination of the lensed quasar at the 1 to 2 mas level, but after applying the correction to the PSF-R, the lensed quasar positions agree with the results from PSF-CS. 
We conclude that the position of the lensed quasars can be robustly determined with precision much better than a milli-arcsecond. 

To compare the precision with the HST, we overlay the results from an analysis of the HST data on this system (Schmidt et al. in prep.) by fitting simultaneously f814W, f475X and f160W bands in \fref{fig:J0659_precision}. 
The precision of the lensed quasar positions is better determined with the AO imaging than the HST imaging, by a factor of 5-10. We stress that this is a comparison of precision. A comparison of accuracy of differential astrometry between the images would require investigating uncertainties in astrometric distortion corrections, which is beyond the scope of this paper.

\section{Astrometric requirement for $H_{0}$: AO v.s. HST}
\label{sec:H0_AOHST}
\citet{BirrerTreu19} show the astrometric requirement for $H_{0}$ given $\Ddt$ in five different possible scenarios for the lens systems. 
In this paper, we focus on the requirement given the measurement of $\Dd$ since $\Dd$ provides the main information on constraining $H_{0}$ under the mass-sheet transformation \citep{GChenEtal20}, noting however that the requirement are the same as discussed above. 
The main factor determining the precision of the cosmological inference then becomes the stellar kinematics since the lens potential can be constrained sufficiently by the extended imaging and line-of-sight mass distribution does not contribute to $H_{0}$ determined from $\Dd$. 
In \tref{tab:astrometricH0}, We use the same five examples of image separations, $\theta_{\rm AB}$, and time delays, motivated by 
\citet{BirrerTreu19}, to examine the astrometric requirements given the measurement of $\Dd$. We assume a $\Lambda$CDM cosmology with fixed $\Omega_{\rm m}=0.3$. 
In those five scenarios, we consider the most stringent astrometric requirements, i.e. those in the presence of spatially resolved kinematics are available from JWST observations. In this case, the contribution from astrometric uncertainty should be less than 3 percent on $H_{0}$ (Yildirim et al in prep.), in order to be subdominant. 

The requirements are expressed as uncertainties in the source position. The image plane astrometric uncertainty can be obtained by $\sigma_{\beta}/\sigma_{\theta}\sim10^{-1}$ under fixed lensing potential, and by $\sigma_{\beta}/\sigma_{\theta}\sim1$ when the positional information is used to determine the lens model (i.e., no extended arc information) \citep{BirrerTreu19}.

Scenario 1 is for a typical cluster-scale lens with image separation of 20 arcsec and a time-delay of 1000 d, 
the relative astrometric requirement is 18 mas in the source plane to not exceed a 3 per cent uncertainty in $H_{0}$. 
This can be achieved by AO imaging. 

Scenario 2 is similar to RXJ1131-1231
\citep{SuyuEtal13,SuyuEtal14,BirrerEtal16,GChenEtal19} 
and B1608+656 \citep{SuyuEtal10}, or for the doubly lensed quasar SDSSJ1206+4332 \citep{BirrerEtal19}. The relative astrometric requirement is 12 mas in the source plane to not exceed a 3 per cent uncertainty in $H_{0}$. This can be also achieved by AO imaging. 

Scenario 3 is smaller separation images of 2 arcsec with a relative time-delay of 10 d. This is similar to HE 0435-1223 \citep{WongEtal17,GChenEtal19} and PG1115+080 \citep{GChenEtal19}. The relative astrometric requirement is 1.8 mas in the source plane to not exceed a 3 per cent uncertainty in $H_{0}$. This can be achieved by AO imaging.

Scenario 4 is the lens system with short time-delays with a relative delay of 4 day and image separation of 1 arcsec. The relative astrometric requirement is 1.5 mas in the source plane to not exceed a 3 per cent uncertainty in $H_{0}$. This can be achieved by AO imaging. 

This last scenario with image separation of 1 arcsec and a relative delay of 1 d is motivated by the lensed supernova iPTF16geu \citep{GoobarEtal17}. The relative astrometric requirement is 0.36 mas in the source plane to not exceed a 3 per cent uncertainty in $H_{0}$.
However, in reality this type of lens system is unlikely to be chosen as the target for the purpose of time-delay cosmography given its short delays. Nevertheless, the astrometric requirement can still be met by AO imaging. 

In sum, AO imaging can meet the astrometric requirements for all kinds of possible scenarios, even in the most stringent circumstances.

\begin{table*}
\caption{Astrometric requirements for five different scenarios of image separations, and time delays at lens redshift, $z_{\rm d}=0.5$, source redshift, $z_{\rm s}=2$ under the mass-sheet transformation. 
The requirements are listed for uncertainty on $H_{0}$ from astrometric uncertainty to be less than the uncertainty on $H_{0}$ from the time-delay uncertainty, $\sigma_{\Delta t}$, and for two scenarios of kinematic data quality. First, we list the requirement for 10 per cent error on H$_0$ from astrometry, which is the typical uncertainty of single-aperture velocity dispersion measurement. Second, we list the requirement for 3 per cent, which is expected from JWST spatially-resolved kinematics (Yildirim et al. in prep). 
The requirements are expressed as uncertainties in the source position. The image plane astrometric uncertainty can be obtained by $\sigma_{\beta}/\sigma_{\theta}\sim10^{-1}$ under fixed lensing potential, and by $\sigma_{\beta}/\sigma_{\theta}\sim1$ when the positional information is used to determine the lens model (i.e., no extended arc information) \citep{BirrerTreu19}. 
Given the precision which can be achieved by the AO imaging data, AO imaging is sufficient for time-delay cosmography in all scenarios listed here.} 
\label{tab:astrometricH0} 
\begin{center}
\begin{tabular}{ lllllll } 
 \hline
 Scenarios & $\theta_{\rm AB}$(arcsec) & $\Delta t_{\rm AB}$ (d) & $\sigma_{\Delta t}$ (d) & $\sigma_{H_{0}}(\sigma_{\beta})\le \sigma_{\sigma_{\Delta t}}$(mas)&$\sigma_{H_{0}}(\sigma_{\beta})\le $ 10 per cent (mas)&$\sigma_{H_{0}}(\sigma_{\beta})\le $ 3 per cent (mas)\\ 
 \hline
 1 & 20& 1000 & 30   & $\sigma_{\beta}=$18 &$\sigma_{\beta}=$60 &$\sigma_{\beta}=$18 \\ 
 2 & 3 & 100  & 3    & $\sigma_{\beta}=$12 &$\sigma_{\beta}=$40 &$\sigma_{\beta}=$12 \\ 
 3 & 2 & 10   & 1    & $\sigma_{\beta}=$6  &$\sigma_{\beta}=$6  &$\sigma_{\beta}=$1.8 \\ 
 4 & 1 & 4    & 0.25 & $\sigma_{\beta}=$3  &$\sigma_{\beta}=$4.8& $\sigma_{\beta}=$1.44 \\ 
 5 & 1 & 1    & 0.025 & $\sigma_{\beta}=$0.3  &$\sigma_{\beta}=$1.2& $\sigma_{\beta}=$0.36 \\ 
 \hline
\end{tabular}
\end{center}
\end{table*}

\section{Conclusions}
\label{sec:summary}
We analyze adaptive optics images of the strong lens J\,0659+1629 obtained with the W.M.Keck Observatory using the laser guide star adaptive optics system to examine the astrometric requirements for time-delay cosmography under the mass-sheet transformation. 
We show that by combining two techniques of PSF reconstruction (PSF-R from telemetry and PSF-CS from the data themselves), we can (1) reconstruct both the core and wings of the AO PSF, (2) subtract the point-like multiple images with residuals consistent with noise, and (3) obtain astrometric precision of$\sim0.3$mas in the source plane which is more then sufficient to meet the requirements even in the most stringent cases, with JWST spatially-resolved kinematics data are available. Therefore, we demonstrate that by applying our techniques to AO data, the astrometric precision will always be the subdominant term in the H$_0$ error budget. 

\section*{Acknowledgements}
The data presented herein were obtained at the W. M. Keck Observatory, which is operated as a scientific partnership among the California Institute of Technology, the University of California and the National Aeronautics and Space Administration. The Observatory was made possible by the generous financial support of the W. M. Keck Foundation. The authors wish to recognize and acknowledge the very significant cultural role and reverence that the summit of Maunakea has always had within the indigenous Hawaiian community.  We are most fortunate to have the opportunity to conduct observations from this mountain. 
We acknowledge support by the National Science Foundation through grant NSF-AST-1906976 and NSF-AST-1907396 "Collaborative Research: Toward a 1\% measurement of the Hubble Constant with gravitational time delays", grant NSF-1836016 "Astrophysics enabled by Keck All Sky Precision Adaptive Optics". We also acknowledge support by the Gordon and Betty Moore Foundation Grant 8548 "Cosmology via Strongly lensed quasars with KAPA".
S.H.S. thanks the Max Planck Society for support through the Max Planck Research Group. This research is supported in part by the Excellence Cluster ORIGINS which is funded by the Deutsche Forschungsgemeinschaft (DFG, German Research Foundation) under Germany’s Excellence Strategy – EXC-2094 – 390783311.




\bibliographystyle{mnras}
\bibliography{AO_cosmography} 

\begin{thebibliography}{}
\makeatletter
\relax
\def\mn@urlcharsother{\let\do\@makeother \do\$\do\&\do\#\do\^\do\_\do\%\do\~}
\def\mn@doi{\begingroup\mn@urlcharsother \@ifnextchar [ {\mn@doi@}
  {\mn@doi@[]}}
\def\mn@doi@[#1]#2{\def\@tempa{#1}\ifx\@tempa\@empty \href
  {http://dx.doi.org/#2} {doi:#2}\else \href {http://dx.doi.org/#2} {#1}\fi
  \endgroup}
\def\mn@eprint#1#2{\mn@eprint@#1:#2::\@nil}
\def\mn@eprint@arXiv#1{\href {http://arxiv.org/abs/#1} {{\tt arXiv:#1}}}
\def\mn@eprint@dblp#1{\href {http://dblp.uni-trier.de/rec/bibtex/#1.xml}
  {dblp:#1}}
\def\mn@eprint@#1:#2:#3:#4\@nil{\def\@tempa {#1}\def\@tempb {#2}\def\@tempc
  {#3}\ifx \@tempc \@empty \let \@tempc \@tempb \let \@tempb \@tempa \fi \ifx
  \@tempb \@empty \def\@tempb {arXiv}\fi \@ifundefined
  {mn@eprint@\@tempb}{\@tempb:\@tempc}{\expandafter \expandafter \csname
  mn@eprint@\@tempb\endcsname \expandafter{\@tempc}}}

\bibitem[\protect\citeauthoryear{{Beckers}}{{Beckers}}{1993}]{Beckers93}
{Beckers} J.~M.,  1993, \mn@doi [\araa] {10.1146/annurev.aa.31.090193.000305},
  \href {http://adsabs.harvard.edu/abs/1993ARA%26A..31...13B} {31, 13}

\bibitem[\protect\citeauthoryear{{Birrer} \& {Treu}}{{Birrer} \&
  {Treu}}{2019}]{BirrerTreu19}
{Birrer} S.,  {Treu} T.,  2019, \mn@doi [\mnras] {10.1093/mnras/stz2254}, \href
  {https://ui.adsabs.harvard.edu/abs/2019MNRAS.489.2097B} {489, 2097}

\bibitem[\protect\citeauthoryear{{Birrer}, {Amara}  \& {Refregier}}{{Birrer}
  et~al.}{2016}]{BirrerEtal16}
{Birrer} S.,  {Amara} A.,   {Refregier} A.,  2016, \mn@doi [\jcap]
  {10.1088/1475-7516/2016/08/020}, \href
  {http://adsabs.harvard.edu/abs/2016JCAP...08..020B} {8, 020}

\bibitem[\protect\citeauthoryear{{Birrer} et~al.,}{{Birrer}
  et~al.}{2019}]{BirrerEtal19}
{Birrer} S.,  et~al., 2019, \mn@doi [\mnras] {10.1093/mnras/stz200}, \href
  {http://adsabs.harvard.edu/abs/2019MNRAS.484.4726B} {484, 4726}

\bibitem[\protect\citeauthoryear{{Birrer} et~al.,}{{Birrer}
  et~al.}{2020}]{BirrerEtal20}
{Birrer} S.,  et~al., 2020, arXiv e-prints, \href
  {https://ui.adsabs.harvard.edu/abs/2020arXiv200702941B} {p. arXiv:2007.02941}

\bibitem[\protect\citeauthoryear{{Chen} et~al.,}{{Chen}
  et~al.}{2016}]{GChenEtal16}
{Chen} G.~C.-F.,  et~al., 2016, \mn@doi [\mnras] {10.1093/mnras/stw991}, \href
  {http://adsabs.harvard.edu/abs/2016MNRAS.462.3457C} {462, 3457}

\bibitem[\protect\citeauthoryear{{Chen} et~al.,}{{Chen}
  et~al.}{2019}]{GChenEtal19}
{Chen} G. C.~F.,  et~al., 2019, \mn@doi [\mnras] {10.1093/mnras/stz2547}, \href
  {https://ui.adsabs.harvard.edu/abs/2019MNRAS.tmp.2193C} {p.~2193}

\bibitem[\protect\citeauthoryear{{Chen}, {Fassnacht}, {Suyu},
  {Y{\i}ld{\i}r{\i}m}, {Komatsu}  \& {Bernal}}{{Chen}
  et~al.}{2020}]{GChenEtal20}
{Chen} G. C.~F.,  {Fassnacht} C.~D.,  {Suyu} S.~H.,  {Y{\i}ld{\i}r{\i}m} A.,
  {Komatsu} E.,   {Bernal} J.~L.,  2020, arXiv e-prints, \href
  {https://ui.adsabs.harvard.edu/abs/2020arXiv201106002C} {p. arXiv:2011.06002}

\bibitem[\protect\citeauthoryear{{Collett} et~al.,}{{Collett}
  et~al.}{2013}]{CollettEtal13}
{Collett} T.~E.,  et~al., 2013, \mn@doi [\mnras] {10.1093/mnras/stt504}, \href
  {http://adsabs.harvard.edu/abs/2013MNRAS.432..679C} {432, 679}

\bibitem[\protect\citeauthoryear{{Falco}, {Gorenstein}  \& {Shapiro}}{{Falco}
  et~al.}{1985}]{FalcoEtal85}
{Falco} E.~E.,  {Gorenstein} M.~V.,   {Shapiro} I.~I.,  1985, \mn@doi [\apjl]
  {10.1086/184422}, \href {http://adsabs.harvard.edu/abs/1985ApJ...289L...1F}
  {289, L1}

\bibitem[\protect\citeauthoryear{{Fassnacht}, {Xanthopoulos}, {Koopmans}  \&
  {Rusin}}{{Fassnacht} et~al.}{2002}]{FassnachtEtal02}
{Fassnacht} C.~D.,  {Xanthopoulos} E.,  {Koopmans} L.~V.~E.,   {Rusin} D.,
  2002, \mn@doi [\apj] {10.1086/344368}, \href
  {http://adsabs.harvard.edu/cgi-bin/nph-bib_query?bibcode=2002ApJ...5
  81..823F&db_key=AST} {581, 823}

\bibitem[\protect\citeauthoryear{{Fassnacht}, {Koopmans}  \&
  {Wong}}{{Fassnacht} et~al.}{2011}]{FassnachtEtal11}
{Fassnacht} C.~D.,  {Koopmans} L.~V.~E.,   {Wong} K.~C.,  2011, \mn@doi
  [\mnras] {10.1111/j.1365-2966.2010.17591.x}, \href
  {http://adsabs.harvard.edu/abs/2011MNRAS.410.2167F} {410, 2167}

\bibitem[\protect\citeauthoryear{{Goobar} et~al.,}{{Goobar}
  et~al.}{2017}]{GoobarEtal17}
{Goobar} A.,  et~al., 2017, \mn@doi [Science] {10.1126/science.aal2729}, \href
  {https://ui.adsabs.harvard.edu/abs/2017Sci...356..291G} {356, 291}

\bibitem[\protect\citeauthoryear{{Gorenstein}, {Falco}  \&
  {Shapiro}}{{Gorenstein} et~al.}{1988}]{GorensteinEtal88}
{Gorenstein} M.~V.,  {Falco} E.~E.,   {Shapiro} I.~I.,  1988, \mn@doi [\apj]
  {10.1086/166226}, \href
  {https://ui.adsabs.harvard.edu/abs/1988ApJ...327..693G} {327, 693}

\bibitem[\protect\citeauthoryear{{Greene} et~al.,}{{Greene}
  et~al.}{2013}]{GreeneEtal13}
{Greene} Z.~S.,  et~al., 2013, \mn@doi [\apj] {10.1088/0004-637X/768/1/39},
  \href {http://adsabs.harvard.edu/abs/2013ApJ...768...39G} {768, 39}

\bibitem[\protect\citeauthoryear{{Hilbert}, {White}, {Hartlap}  \&
  {Schneider}}{{Hilbert} et~al.}{2007}]{HilbertEtal07}
{Hilbert} S.,  {White} S.~D.~M.,  {Hartlap} J.,   {Schneider} P.,  2007,
  \mn@doi [\mnras] {10.1111/j.1365-2966.2007.12391.x}, \href
  {http://adsabs.harvard.edu/abs/2007MNRAS.382..121H} {382, 121}

\bibitem[\protect\citeauthoryear{{Hilbert}, {Hartlap}, {White}  \&
  {Schneider}}{{Hilbert} et~al.}{2009}]{HilbertEtal09}
{Hilbert} S.,  {Hartlap} J.,  {White} S.~D.~M.,   {Schneider} P.,  2009,
  \mn@doi [\aap] {10.1051/0004-6361/200811054}, \href
  {http://adsabs.harvard.edu/abs/2009A%26A...499...31H} {499, 31}

\bibitem[\protect\citeauthoryear{{Jee}, {Komatsu}  \& {Suyu}}{{Jee}
  et~al.}{2015}]{JeeEtal15}
{Jee} I.,  {Komatsu} E.,   {Suyu} S.~H.,  2015, \mn@doi [\jcap]
  {10.1088/1475-7516/2015/11/033}, \href
  {http://adsabs.harvard.edu/abs/2015JCAP...11..033J} {11, 033}

\bibitem[\protect\citeauthoryear{{Kochanek}}{{Kochanek}}{2020}]{Kochanek20}
{Kochanek} C.~S.,  2020, \mn@doi [\mnras] {10.1093/mnras/staa344}, \href
  {https://ui.adsabs.harvard.edu/abs/2020MNRAS.493.1725K} {493, 1725}

\bibitem[\protect\citeauthoryear{{Kochanek}}{{Kochanek}}{2021}]{Kochanek21}
{Kochanek} C.~S.,  2021, \mn@doi [\mnras] {10.1093/mnras/staa4033}, \href
  {https://ui.adsabs.harvard.edu/abs/2021MNRAS.501.5021K} {501, 5021}

\bibitem[\protect\citeauthoryear{{Kochanek}, {Keeton}  \& {McLeod}}{{Kochanek}
  et~al.}{2001}]{KochanekEtal01}
{Kochanek} C.~S.,  {Keeton} C.~R.,   {McLeod} B.~A.,  2001, \mn@doi [\apj]
  {10.1086/318350}, \href {http://adsabs.harvard.edu/abs/2001ApJ...547...50K}
  {547, 50}

\bibitem[\protect\citeauthoryear{{Koopmans} et~al.,}{{Koopmans}
  et~al.}{2003}]{KoopmansEtal03b}
{Koopmans} L.~V.~E.,  et~al., 2003, \mn@doi [\apj] {10.1086/377434}, \href
  {http://adsabs.harvard.edu/abs/2003ApJ...595..712K} {595, 712}

\bibitem[\protect\citeauthoryear{{Paraficz} \& {Hjorth}}{{Paraficz} \&
  {Hjorth}}{2009}]{ParaficzHjorth09}
{Paraficz} D.,  {Hjorth} J.,  2009, \mn@doi [\aap]
  {10.1051/0004-6361/200913307}, \href
  {http://adsabs.harvard.edu/abs/2009A%26A...507L..49P} {507, L49}

\bibitem[\protect\citeauthoryear{Ragland}{Ragland}{2018}]{Ragland18}
Ragland S.,  2018, {A novel technique to measure residual systematic segment
  piston errors of large aperture optical telescopes}.
~ Vol. 10700, SPIE, \mn@doi{10.1117/12.2313017}, \url
  {https://doi.org/10.1117/12.2313017}

\bibitem[\protect\citeauthoryear{{Ragland} et~al.,}{{Ragland}
  et~al.}{2018}]{RaglandEtal18}
{Ragland} S.,  et~al., 2018, {Status of point spread function determination for
  Keck adaptive optics}.
 Society of Photo-Optical Instrumentation Engineers (SPIE) Conference Series
  Vol. 10703, \mn@doi{10.1117/12.2312975, }

\bibitem[\protect\citeauthoryear{{Refsdal}}{{Refsdal}}{1964}]{Refsdal64}
{Refsdal} S.,  1964, \mnras, \href
  {http://adsabs.harvard.edu/cgi-bin/nph-bib_query?bibcode=1964MNRAS.128..307R&db_key=AST}
  {128, 307}

\bibitem[\protect\citeauthoryear{{Riess}, {Casertano}, {Yuan}, {Macri}  \&
  {Scolnic}}{{Riess} et~al.}{2019}]{RiessEtal19}
{Riess} A.~G.,  {Casertano} S.,  {Yuan} W.,  {Macri} L.~M.,   {Scolnic} D.,
  2019, arXiv e-prints, \href
  {https://ui.adsabs.harvard.edu/\#abs/2019arXiv190307603R} {p.
  arXiv:1903.07603}

\bibitem[\protect\citeauthoryear{{Rousset}, {Fontanella}, {Kern}, {Gigan}  \&
  {Rigaut}}{{Rousset} et~al.}{1990}]{RoussetEtal90}
{Rousset} G.,  {Fontanella} J.~C.,  {Kern} P.,  {Gigan} P.,   {Rigaut} F.,
  1990, \aap, \href {http://adsabs.harvard.edu/abs/1990A%26A...230L..29R} {230,
  L29}

\bibitem[\protect\citeauthoryear{{Rusu} et~al.,}{{Rusu}
  et~al.}{2017}]{RusuEtal17}
{Rusu} C.~E.,  et~al., 2017, \mn@doi [\mnras] {10.1093/mnras/stx285}, \href
  {http://adsabs.harvard.edu/abs/2017MNRAS.467.4220R} {467, 4220}

\bibitem[\protect\citeauthoryear{{Rusu} et~al.,}{{Rusu}
  et~al.}{2020}]{RusuEtal20_H0LiCOW}
{Rusu} C.~E.,  et~al., 2020, \mn@doi [\mnras] {10.1093/mnras/stz3451}, \href
  {https://ui.adsabs.harvard.edu/abs/2020MNRAS.498.1440R} {498, 1440}

\bibitem[\protect\citeauthoryear{{Suyu}, {Marshall}, {Blandford}, {Fassnacht},
  {Koopmans}, {McKean}  \& {Treu}}{{Suyu} et~al.}{2009}]{SuyuEtal09}
{Suyu} S.~H.,  {Marshall} P.~J.,  {Blandford} R.~D.,  {Fassnacht} C.~D.,
  {Koopmans} L.~V.~E.,  {McKean} J.~P.,   {Treu} T.,  2009, \mn@doi [\apj]
  {10.1088/0004-637X/691/1/277}, \href
  {http://adsabs.harvard.edu/abs/2009ApJ...691..277S} {691, 277}

\bibitem[\protect\citeauthoryear{{Suyu}, {Marshall}, {Auger}, {Hilbert},
  {Blandford}, {Koopmans}, {Fassnacht}  \& {Treu}}{{Suyu}
  et~al.}{2010}]{SuyuEtal10}
{Suyu} S.~H.,  {Marshall} P.~J.,  {Auger} M.~W.,  {Hilbert} S.,  {Blandford}
  R.~D.,  {Koopmans} L.~V.~E.,  {Fassnacht} C.~D.,   {Treu} T.,  2010, \mn@doi
  [\apj] {10.1088/0004-637X/711/1/201}, \href
  {http://adsabs.harvard.edu/abs/2010ApJ...711..201S} {711, 201}

\bibitem[\protect\citeauthoryear{{Suyu} et~al.,}{{Suyu}
  et~al.}{2013}]{SuyuEtal13}
{Suyu} S.~H.,  et~al., 2013, \mn@doi [\apj] {10.1088/0004-637X/766/2/70}, \href
  {http://adsabs.harvard.edu/abs/2013ApJ...766...70S} {766, 70}

\bibitem[\protect\citeauthoryear{{Suyu} et~al.,}{{Suyu}
  et~al.}{2014}]{SuyuEtal14}
{Suyu} S.~H.,  et~al., 2014, \mn@doi [\apjl] {10.1088/2041-8205/788/2/L35},
  \href {http://adsabs.harvard.edu/abs/2014ApJ...788L..35S} {788, L35}

\bibitem[\protect\citeauthoryear{{Tihhonova} et~al.,}{{Tihhonova}
  et~al.}{2018}]{TihhonovaEtal18}
{Tihhonova} O.,  et~al., 2018, \mn@doi [\mnras] {10.1093/mnras/sty1040}, \href
  {https://ui.adsabs.harvard.edu/abs/2018MNRAS.477.5657T} {477, 5657}

\bibitem[\protect\citeauthoryear{{Tihhonova} et~al.,}{{Tihhonova}
  et~al.}{2020}]{TihhonovaEtal20}
{Tihhonova} O.,  et~al., 2020, \mn@doi [\mnras] {10.1093/mnras/staa1436}, \href
  {https://ui.adsabs.harvard.edu/abs/2020MNRAS.498.1406T} {498, 1406}

\bibitem[\protect\citeauthoryear{{Treu} \& {Koopmans}}{{Treu} \&
  {Koopmans}}{2002}]{TreuKoopmans02}
{Treu} T.,  {Koopmans} L.~V.~E.,  2002, \mn@doi [\mnras]
  {10.1046/j.1365-8711.2002.06107.x}, \href
  {http://adsabs.harvard.edu/abs/2002MNRAS.337L...6T} {337, L6}

\bibitem[\protect\citeauthoryear{Watson}{Watson}{1997}]{Watson97}
Watson J.,  1997, in WESCON/97 Conference Proceedings. pp 490--494,
  \mn@doi{10.1109/WESCON.1997.632376}

\bibitem[\protect\citeauthoryear{Wizinowich et~al.,}{Wizinowich
  et~al.}{2006}]{wizinowich06}
Wizinowich P.~L.,  et~al., 2006, Publications of the Astronomical Society of
  the Pacific, 118, 297

\bibitem[\protect\citeauthoryear{{Wong} et~al.,}{{Wong}
  et~al.}{2017}]{WongEtal17}
{Wong} K.~C.,  et~al., 2017, \mn@doi [\mnras] {10.1093/mnras/stw3077}, \href
  {http://adsabs.harvard.edu/abs/2017MNRAS.465.4895W} {465, 4895}

\bibitem[\protect\citeauthoryear{{Y{\i}ld{\i}r{\i}m}, {Suyu}  \&
  {Halkola}}{{Y{\i}ld{\i}r{\i}m} et~al.}{2020}]{YidrimEtal20}
{Y{\i}ld{\i}r{\i}m} A.,  {Suyu} S.~H.,   {Halkola} A.,  2020, \mn@doi [\mnras]
  {10.1093/mnras/staa498}, \href
  {https://ui.adsabs.harvard.edu/abs/2020MNRAS.493.4783Y} {493, 4783}

\makeatother
\end{thebibliography}




\appendix
\section{The error breakdowns of the PSF-R for J\,0659+1629}
\label{appexdix1}
We list the error breakdowns of the PSF-R for J\,0659+1629 AO imaging observation in the following:
\begin{itemize}
\item Fitting error:  165 nm
\item Aliasing error:  69 nm
\item TT residuals:  354 nm
\item DM residuals:  264 nm
\item Focal anisoplanatism:  187 nm
\item Static Aberration:  230 nm
\item The total wavefront error is  561 nm.
\end{itemize}
Note that the relatively large static aberration (230 nm) comes from an issue with telescope phasing that was addressed the night after our observations. This could be responsible for the poor reconstruction of the core of the PSF-R.


\bsp	
\label{lastpage}
\end{document}